\documentclass[twocolumn]{aastex631}

\let\tablenum\relax
\usepackage{dirtytalk}
\usepackage{graphicx}
\usepackage{float}
\usepackage{enumitem}
\usepackage{CJK}
\usepackage{siunitx}
\usepackage{siunitx}
\usepackage{multirow}
\usepackage{comment}
\usepackage{longtable}
\usepackage{tabularx}

\begin{document}
\begin{CJK*}{UTF8}{gbsn}

\title{Disk-Locking Regulates Stellar Rotation in Young Clusters: Insights from NGC 2264}
 
\correspondingauthor{Chengyuan Li}
\email{lichengy5@mail.sysu.edu.cn}

\author[0009-0001-7230-3348]{Yutian Bu (卜玉田)}
\affiliation{School of Physics and Astronomy, Sun Yat-sen University, Zhuhai, 519082, People's Republic of China}
\affiliation{CSST Science Center for the Guangdong-Hong Kong-Macau Greater Bay Area, Zhuhai, 519082, People's Republic of China}

\author[0000-0001-9131-6956]{Chenyu He (贺辰昱)}
\affiliation{School of Physics and Astronomy, Sun Yat-sen University, Zhuhai, 519082, People's Republic of China}
\affiliation{CSST Science Center for the Guangdong-Hong Kong-Macau Greater Bay Area, Zhuhai, 519082, People's Republic of China}

\author[0000-0001-8060-1321]{Min Fang (房敏)}
\affiliation{Purple Mountain Observatory, Chinese Academy of Sciences, 10 Yuanhua Road, Nanjing 210033, People's
Republic of China}

\author[0000-0002-3084-5157]{Chengyuan Li (李程远)}
\affiliation{School of Physics and Astronomy, Sun Yat-sen University, Zhuhai, 519082, People's Republic of China}
\affiliation{CSST Science Center for the Guangdong-Hong Kong-Macau Greater Bay Area, Zhuhai, 519082, People's Republic of China}

\begin{abstract}
Many young clusters possess extended main sequences, a phenomenon commonly ascribed to stellar rotation. However, the mechanism behind their very wide stellar rotation distributions remains unclear. A proposed explanation is that magnetic star-disk interaction can regulate stellar rotation, i.e., proto-stars with longer disk lifetimes will eventually evolve into slow rotators, and vice versa. To examine this hypothesis, we took the star forming region, NGC 2264, as a test bed. We have studied its high-mass pre-main-sequence and zero-age main-sequence stars. We found that on average, disk-less pre-main-sequence stars rotate faster than their disk-bearing counterparts. The stellar rotation distribution of its zero-age main-sequence stars is similar to evolved young clusters. We conclude that disk-locking may play a crucial role in the rotational velocity distribution of intermediate-mass early-type stars. We suggest that the observed wide stellar rotation distribution in many young clusters can occur in their early stages. 

\end{abstract}
\keywords{stars: rotation --- open clusters and associations:
  individual: NGC 2264 --- galaxies: star clusters: general.}



\section{Introduction}
Most open clusters younger than $\sim$ \SI{2}{Gyr} in the local group display extended main-sequences (eMSs) \citep[e.g., ][]{2009A&A...497..755M,2018ApJ...869..139C}. Among them, many intermediate-age clusters (1--2 Gyrs) exhibit extended main-sequence turnoffs (eMSTOs) and their younger counterparts ($<\sim$\SI{600}{Myr}) exhibit extended or even split main-sequences (MSs) \citep[e.g.,][]{2015MNRAS.450.3750M,2019ApJ...883..182S,2022ApJ...938...42H}. It is widely accepted that rapidly rotating stars significantly contribute to the formation of eMSs  \citep[e.g.,][]{2019ApJ...876..113S,2019ApJ...883..182S}, through the effects of rotational gravitational darkening \citep[e.g.,][]{1968ApJ...151..203F}, rotational mixing \citep[e.g.,][]{2013ApJ...776..112Y} and rotation-induced self-extinguishing stellar disks \citep[e.g.,][]{2023MNRAS.521.4462D}. Since rapid rotation is a common property of early-type (earlier than F-type) stars, the rotation distributions of clusters typically do not exhibit a Gaussian-like form, however. Their extended or bifurcated main sequences suggest a very wide or bimodal distribution of stellar rotations \citep[e.g.,][]{2018MNRAS.477.2640M,2018AJ....156..116M}. Since the majority of stars are formed in clusters, a bimodal distribution is similarly observed among field stars \citep{2012A&A...537A.120Z,2021ApJ...921..145S}, as anticipated. While the rotational velocity distribution in clusters is more extensively spread than that of field stars \citep{2021ApJ...921..145S}, the mass range for stars exhibiting a bimodal rotation distribution also varies between clusters and field stars \citep[e.g.,][]{2018MNRAS.477.2640M,2020MNRAS.492.2177K}.

The mechanism driving the wide variations in stellar rotations remains an unanswered question, i.e., how a significant number of slowly or non-rotating stars among early-type stars is formed has not been addressed yet. Numerous scenarios have been proposed to explain the variations in stellar rotation. \cite{2022NatAs...6..480W} found that binary mergers can reproduce the observed morphology of the split MSs in many Magellanic Cloud clusters (hereafter binary-merger model). They suggest that most slowly rotating stars have experienced a merger event with a companion, indicating that these objects are blue straggler stars (BSSs). Their model is based on two fundamental assumptions. First, most BSSs that have experienced mergers are magnetars \citep{2019Natur.574..211S}, and the magnetic braking mechanism enables them to rapidly dissipate angular momentum, leading to the formation of slow rotators. Secondly, the interaction between binary stars and circum-binary matter (such as circum-binary disks) can accelerate the merger process. Alternatively, a significant proportion of high-order multiple systems within star clusters may facilitate mergers through Lidov-Kozai cycles or gravitational encounters with passing stars. These mechanisms may account for the formation of BSSs in young clusters, aligning with the observed blue sequence stars. 

Another scenario suggests that tidal interactions between the components of close binaries can effectively reduce stellar rotation \citep[][hereafter binary-locking model]{2015MNRAS.453.2637D,2017NatAs...1E.186D}. Their model implies an assumption that binary systems undergoing tidal slowing must have a small mass ratio, as high mass ratio binaries would reside at the redder end of the main sequence, being unresolved by telescopes. Although subsequent observations have revealed that most slowly rotating early-type stars of open clusters are not short-period binaries \citep{2023MNRAS.525.5880H,2024ApJ...968...22B}, and numerical simulations indicate that only those binaries with mass ratios close to 1 could be dynamically locked within the limited age of the cluster \citep{2023ApJ...949...53W}, the presence of tidally locked binaries with periods ranging from 100 to 500 days lends potential support to their model \citep{2004ApJ...616..562A}. 

Another model proposed that the bimodal velocity distribution is set during cluster formation or early evolution stages \citep[][hereafter disc-locking model]{2020MNRAS.495.1978B}. Furthermore, they suggested that a bimodal spin rate distribution could persist during MS stages and the corresponding stellar masses may extend up to $\sim$ \SI{5}{M_{\odot}}. In their model, surface velocities of three PMS stars, with different initial periods and masses corresponding to MSTO age of $\sim$ \SI{1.5}{Gyr}, evolve to $\sim$ \SIrange{60}{230}{\kilo\meter\per\second} at the arrival on the MS. They consider ``disk-locking'' as the mechanism differentiate the velocities of the PMS stars. In other words, due to the magnetic star-disk interaction, stars with longer disk lifetimes will become slow rotators on the MS and stars clearing their disks earlier will evolve to fast rotators. A more comprehensive discussion about these scenarios can be found in the review article by \cite{LI2024} and references therein. 

The binary-locking model indicates that unless a substantial number of very close binaries existed at the cluster's initial formation or some binaries were synchronised at birth, most binaries in star-forming clusters did not have enough time to achieve tidal locking. Consequently, most of their early-type zero-age main-sequence (ZAMS) stars should be rapidly rotating, while both the binary-merger model and the disk-locking model can indicate a broad range of rotational velocities for early-type ZAMS stars. The distinction between the binary-merger model and the disk-locking model lies in the influence of the disk on stellar rotation. Concerning the binary-merger model, 
The loss of angular momentum can occur through the ejection of the circum-binary disk or interactions with a third body, resulting in the merger of the binary system (thus the formation of a magnetic BSS according to \cite{2022NatAs...6..480W}). The overall effect is that disk-less stars would tend to be slow rotators. These disk-less stars may occupy a fraction of $\sim$25\% to 50\% of the entire population \citep[e.g.,][]{2016MNRAS.458.4368M,2017ApJ...834..156L,2019ApJ...883..182S}.

In contrast, the disk-locking model posits that stellar systems that lose their disks by external forces, such as photoionization, at an early stage, lacking continuous magnetic disk-locking, will retain their initial higher angular momentum, leading them to evolve into fast-rotating MS stars. Therefore, investigating the rotation of protostars in star-forming clusters and their relationship with the presence of protostellar disks will be instrumental in distinguishing between these models. In summary, to test the three models through observations, it is crucial to study very young star clusters (star-forming clusters, ≤10 Myr). 

In some young clusters, low mass ($<\sim$\SI{2}{M_{\odot}}) PMS stars display bimodal period distributions, and disk-bearing objects rotate more slowly on average than sources without circumstellar disks, indicating magnetic star-disk interaction is sufficient to slow the rotation of low-mass stars, such as NGC 2264 \citep[e.g.,][]{2007ApJ...671..605C,2016A&A...586A..96L,2017A&A...599A..23V}, IC 348 \citep[e.g.,][]{2006ApJ...649..862C}, Upper Scorpius and $\rho$ Ophiuchus \citep{2018AJ....155..196R}, and Upper Centaurus-Lupus and Lower Centaurus-Crux \citep{2022AJ....164...80R}. However, whether disk-locking is suitable for stars resulting eMSs (i.e., stars with masses ranging from $\sim$\SIrange{1.5}{5}{M_\odot} \citep{2018MNRAS.477.2640M}) requires further investigation.

In this work, we took NGC 2264, an extensively studied star formation region with low foreground extinction, as a test bed to investigate if and how disk-locking regulates its initial stellar rotations. Our objectives are to examine: (1) whether the rotational velocity distributions of its disk-bearing and disk-less stars differ, and (2) whether the NGC 2264 ZAMS stars display similar rotation distributions with some evolved open clusters harboring eMSs. Our results indicate that disk-locking does play an important role in rotational velocity distributions, which could be set during cluster formation stages.

This article is organized as follows. In section \ref{sec:methods} and \ref{sec:results}, we present data reduction and main results, respectively. Our discussion and conclusion are displayed in section \ref{sec:discussion} and \ref{sec:conclusion}, respectively.

\section{Methods} \label{sec:methods}
\subsection{Membership Determination \& Isochrone Fitting} \label{sec:membership}

\begin{figure*}
\centering
\begin{tabular}{cc}
    \includegraphics[width=0.9\textwidth]{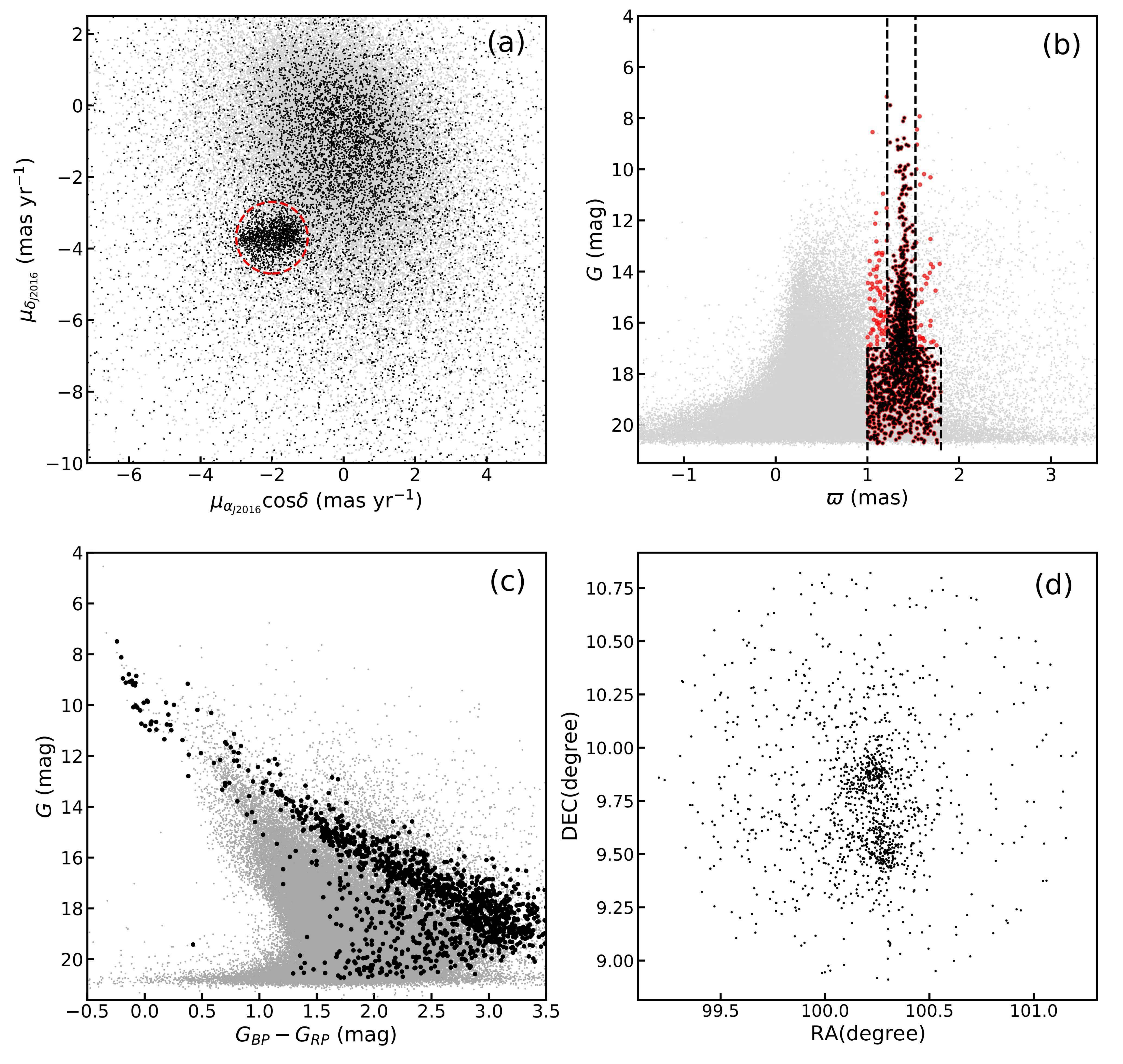}
\end{tabular}
	\caption{(a), the distribution of proper motions (PM) of the stars with $\SI{1.0}{mas}<\varpi<\SI{1.8}{mas}$ (black dots). The dashed circle shows the PM range of the member candidates which is centered at $\mathrm{(-2.0,3.7)\,mas\,yr^{-1}}$ with a radius of $\SI{1}{mas}$. (b) the $\varpi$ distributions of the member candidates (red dots) and selected stars (black dots) of NGC 2264. The dashed lines show the range of the $\varpi$ of the member stars (see the text for details). (c), the CMD of the member stars (black dots). In panels (a), (b), and (c), the grey dots represent the distribution of all stars within $1^{\circ}$ of the cluster center. (d), the spatial distribution of the member stars.}\label{fig:Mselection}
\end{figure*}

We adopted the coordinates of the center of NGC 2264, $\alpha_{J2015.5} = 06\textsuperscript{h} 40\textsuperscript{m} 52\raisebox{0.9ex}{\makebox[1pt][c]{\scriptsize \hspace{0.3em}s}}.1 $ and $\delta_{J2015.5} = \ang[angle-symbol-over-decimal=false]{+9;52;37}$, from \cite{3D_kinematics_2021A&A...647A..19T}. We obtained astrometric and photometric data from Gaia Early Data Release 3 \citep[EDR3;][]{2016A&A...595A...1G,2021A&A...649A...1G} of stars within $\SI{1}{^{\circ}}$ of the cluster center. The radius of the selected field of view corresponds to $\sim$ \SI{13}{pc} for a distance to NGC 2264 of $\sim$ \SI{739}{pc}. We first selected stars whose parallax, $\varpi$, are within $\SI{1.0}{mas}<\varpi<\SI{1.8}{mas}$, whose $\varpi$ are around the average $\varpi$ ($\sim \SI{1.354}{mas}$) of the NGC 2264 member stars reported by \cite{2020A&A...633A..99C}. Figure \ref{fig:Mselection} (a) shows the distribution of their proper motions. The proper motions of NGC 2264 members are well separated from those of the field stars and concentrate around $(\mu_{\alpha}\cos\delta, \mu_{\delta})\approx \mathrm{(-2.0,3.7)\,mas\,yr^{-1}}$. We confined the proper motions of the cluster member stars to have $\sqrt{(\mu_{\alpha}\cos\delta-(-2.0))^2+(\mu_{\delta}-3.7)^2}<\SI{1}{mas\,yr^{-1}}$ (within the dashed red circle in Figure \ref{fig:Mselection}(a)). By this process, we have selected 1522 stars as the candidate member stars of NGC 2264.

We then calculated the mean ($\varpi_{a}$) and standard dispersion ($\sigma_{\varpi}$) of the $\varpi$ of the candidate member stars. As stars with $G<\SI{17}{mag}$ have smaller typical $\varpi$ measurement error than those of fainter stars, we constrained the member stars to have $\varpi_{a}$ within $\varpi_{a}\pm \sigma_{\varpi}$ for the stars with $G<\SI{17}{mag}$, while within $\varpi_{a}\pm
3\sigma_{\varpi}$ for stars with $G>\SI{17}{mag}$, due to the larger typical parallax measurement error for stars with $G>\SI{17}{mag}$ \citep{2021A&A...649A...1G}. We then calculated the $\varpi_{a}$ and $\sigma_{\varpi}$ of the remaining stars. We repeated this process until the differences of $\varpi_{a}$ and
$\sigma_{\varpi}$ between subsequent iterations were smaller than 1\%
of the mean $\varpi$ measurement error of stars with $G<\SI{17}{mag}$. By stellar $\varpi$, we finally selected 1417 member stars from the candidate members for NGC 2264 (see Figure \ref{fig:Mselection} (b)). Their $\varpi_{a}$ and $\sigma_{\varpi}$ are $\SI{1.37}{mas}$ and $\SI{0.15}{mas}$, respectively. The obtained $\varpi_{a}$ is close to the average $\varpi$ reported by \cite{2020A&A...633A..99C}. The average $\mu_{\alpha}\cos\delta$ and $\mu_{\delta}$ is $-1.96\pm0.45$ $\SI{}{mas\,,yr^{-1}}$ and $-3.68\pm0.36$ $\SI{}{mas\,,yr^{-1}}$, respectively. The average $\mu_{\delta}$ is close to the value ($\SI{-3.727}{mas\,,yr^{-1}}$) reported by \cite{2020A&A...633A..99C},  while the average $\mu_{\alpha}\cos\delta$ is not close to the result ($\SI{-1.690}{mas\,yr^{-1}}$) of \cite{2020A&A...633A..99C}. As NGC 2264 is long thought to have several parts \citep[e.g.,][]{2003ARA&A..41...57L, 2020A&A...636A..80B} whose average proper motions are slightly 
different ranging from $\SI{-1.62}{mas\,yr^{-1}}$ to
$\SI{-2.29}{mas\,yr^{-1}}$ \citep{2019ApJ...870...32K}, the difference of the average $\mu_{\alpha}\cos\delta$ of NGC 2264 member stars between this paper and \cite{2020A&A...633A..99C}
is normal. In Figure \ref{fig:Mselection} (c) and (d), we show the CMD of the obtained cluster members and the field stars, and the spatial distribution of the selected member stars respectively. In Figure \ref{fig:Mselection} (d), NGC 2264 appears to have two over-dense regions. The hierarchical structure of NGC 2264 has been reported by many references\citep[e.g.,][]{2003ARA&A..41...57L,2008hsf1.book..966D,2009AJ....138.1116S,2018A&A...609A..10V,2019ApJ...870...32K,2022ApJS..262....7H}, where it has been suggested to have at least two substructures.  Intriguingly, we found that NGC 2264 displays two main-sequence turn-ons (MSTOn, where stars end their PMS stages and join MS) regions in the CMD, indicating an age spread, whether or not these two MSTOns correspond to the two over-dense regions in spatial is yet to be examined, which is beyond the scope of this article. We also compared our selected member stars with those selected by \cite{2022ApJS..262....7H} and \cite{2023A&A...670A..37F}, which are shown in the appendix (Fig. \ref{fig:comparison}). For the member stars provided by \cite{2023A&A...670A..37F}, Fig. \ref{fig:comparison} shows the compromise samples between contamination and sample completeness suggested by \cite{2023A&A...670A..37F} (for details, see \cite{2023A&A...670A..37F}) who were also cross-matched with Gaia EDR3. As Fig. \ref{fig:comparison} shown, our selected members are well consistent those from both references. For stars with $G<\SI{12.5}{mag}$\footnote{The spectroscopic samples observed by Canada-France-Hawaii Telescope in this paper have $G<\SI{12.5}{mag}$.}, $85\%$ and $100\%$ of the member stars selected in this paper are included in those from \cite{2022ApJS..262....7H} and \cite{2023A&A...670A..37F}, respectively.

\begin{figure}
    \centering
    \includegraphics[width=0.47\textwidth]{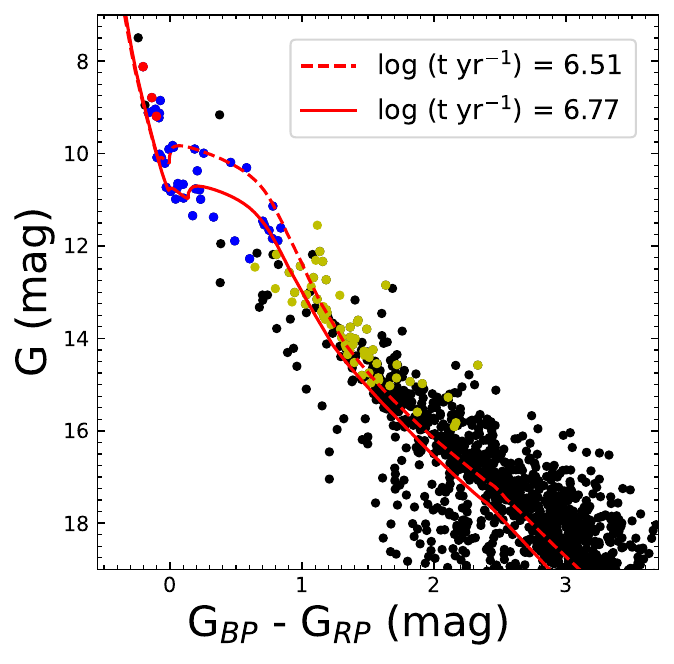}
    \caption{The isochrone-fitted CMD of NGC 2264. Blue circles represent stars observed by CFHT. Red circles represent spectroscopic binaries observed by CFHT. Yellow circles are PMS stars with masses over \SI{1.3}{M_{\odot}}, derived from the catalog provided by \cite{2017A&A...599A..23V}. Black circles represent all other members. The relatively young ($\log (t\ yr^{-1}) = 6.51$) and old ($\log (t\ yr^{-1}) = 6.77$) isochrones are indicated by red dashed and solid lines, respectively.
    }
    \label{fig:members}
\end{figure}

We utilized two isochrones generated from MESA Isochrones \& Stellar Tracks (MIST, version 1.2) \citep{2016ApJ...823..102C} to fit the CMD, our fitting is shown in Fig. \ref{fig:members}. To obtain stellar rotation and examine its relation with disks, we chose 43 relatively massive stars in NGC 2264 to observe, using Canada\text{-}France\text{-}Hawaii Telescope (CFHT), spectroscopic single stars and binaries highlighted as blue and  red dots, respectively, in Fig. \ref{fig:members} (see Section \ref{sec:vsini_cal} for details). We visually determined the best-fitting isochrone parameters and related uncertainties.\footnote{The uncertainties are the steps we used in isochrone-fitting.} The isochrones differ only in age, i.e., $\log (t\ yr^{-1}) = 6.51$ ($\sim$ \SI{3.2}{Myr}) and $\log (t\ yr^{-1}) = 6.77$ ($\sim$ \SI{5.9}{Myr}), both with uncertainties of $\log (t\ yr^{-1}) = 0.1$. Applying an initial rotation velocity of 0.0 critical velocity, these best-fitting isochrones indicate that the cluster has a metallicity of [Fe/H] = $-0.1 \pm \SI{0.1}{dex}$, an extinction of $A_{V} = 0.2 \pm \SI{0.05}{mag}$, and a distance of $700 \pm \SI{20}{pc}$ (distance modulus of $(m-M)_0 \sim 9.2 \pm 0.06 \SI{}{mag}$). These results align with those of \cite{2021MNRAS.504..356D}. They reported an age of $\log (t\ yr^{-1}) = 6.847 \pm 0.051$ ($\sim 7.03 \pm \SI{0.9}{Myr}$), a metallicity of [Fe/H] = $-0.184 \pm \SI{0.11}{dex}$ , an extinction of $A_{V} = 0.404 \pm \SI{0.08}{mag}$ and a distance of $702 \pm \SI{7}{pc}$ for NGC 2264. The cluster has long been found to have an age spread in order of \SI{1}{Myr} \citep[e.g.,][]{2008hsf1.book..966D,2018A&A...609A..10V}, and to embed star-formation regions\citep{2008AJ....135..441S,2009AJ....138.1116S}. The isochrone fitting result in Figure \ref{fig:members} indicates an age spread of $\sim$ \SI{3}{Myr}. Even though it is close to that reported by \cite{2018A&A...609A..10V} (4--\SI{5}{Myr}) obtained by exploring the isochrone ages of the PMS stars in the Hertzsprung–Russell diagram, the age spread we obtained might be polluted by differential extinction.  We note some stars appear below the best-fitting isochrones, which can not be excluded by proper motions or parallax. Their luminosity spread might be driven by star spots or circumstellar disk orientations rather than age spread \citep{2009A&A...504..461F,2021ApJ...908...49F,2013ApJS..207....5F,2013A&A...549A..15F}. These stars do not significantly affect our main conclusion.

\subsection{Spectroscopic Data Reduction}\label{sec:vsini_cal}
The spectra of 43 members (blue and red dots in Fig. \ref{fig:members}) were observed by ESPaDOnS with a resolution of $\sim$68000 and a signal-to-noise ratio of $\sim$30 at $\sim$ \SI{4440}{\AA}, from CFHT program 23BS014. We calculated stellar projected rotational velocities ($v\sin i$) of the sources by fitting observational absorption lines with synthetic spectra, except for 3 spectroscopic binaries (red dots in Fig. \ref{fig:members}).  Since effective temperature of those stars covers a wide range, we utilized Mg \uppercase\expandafter{\romannumeral2} ($\sim$ \SI{4481}{\AA}), Ca \uppercase\expandafter{\romannumeral2} ($\sim$ \SI{8542}{\AA}) and Ca \uppercase\expandafter{\romannumeral2} ($\sim$ \SI{5019}{\AA}) absorption lines, which are nearly unaffected by atmospheric absorption lines or circumstellar disks, to calculate atmospheric temperatures for stars with different stellar types. Synthetic spectra were produced by \cite{2014MNRAS.440.1027C}. We selected effective temperature, $T_\mathrm{eff}$, of the synthetic spectra ranging from \SI{4000}{K} to \SI{21000}{K}, with increments of \SI{250}{K} for $T_\mathrm{eff}$ below \SI{13000}{K} and \SI{1000}{K} for $T_\mathrm{eff}$ above \SI{13000}{K}. The model provides [$\alpha$/Fe] values of 0.0 and 0.4. Considering NGC 2264 is a young open cluster with no evidence of $\alpha$-enrichment, we set [$\alpha$/Fe] to be \SI{0.0}{dex}. The metallicity of the best-fitting isochrones is [Fe/H] = \SI{-0.1}{dex}, but it is not available with [$\alpha$/Fe] = \SI{0.0}{dex} in the model. Therefore, we used model spectra with [Fe/H] = \SI{0.0}{dex}, the closest value allowed by the model. In addition, we applied surface gravity $\log g$ from \SI{3.5}{dex} to \SI{5.0}{dex} in steps of \SI{0.5}{dex}, except when the spectra with $\log g$ = \SI{5.0}{dex} was not available. Further details can be found in \cite{2014MNRAS.440.1027C}.

\begin{figure}
    \centering
    \includegraphics[width=0.47\textwidth]{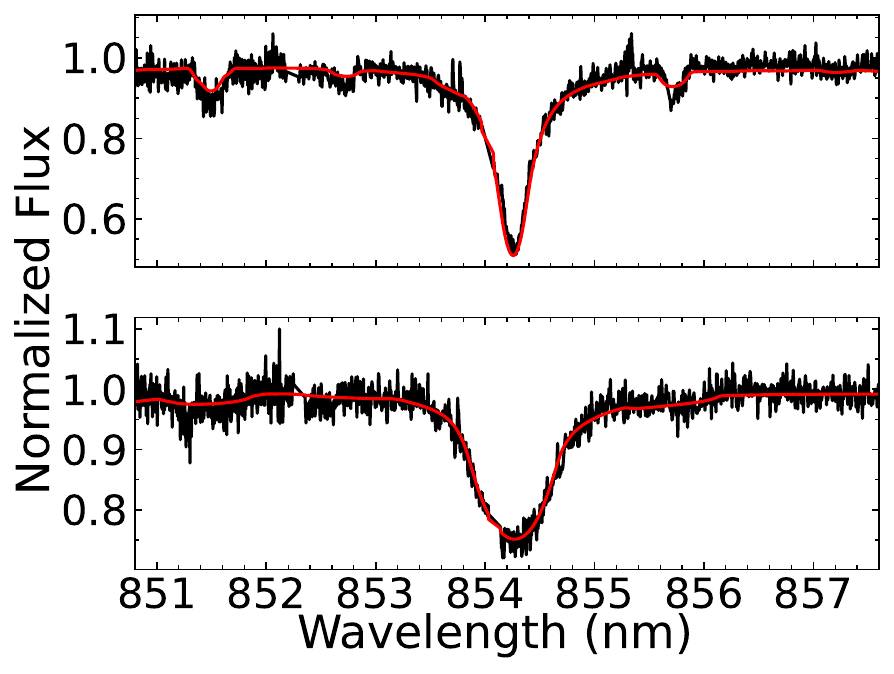}
    \caption{Illustrative fitting results. The upper and lower panels show the observational spectra (black lines) and model fits (red curves) for a slowly rotating star (the disk-bearing star and a rapidly rotating star, respectively.}
    \label{fig:vsini_demo}
\end{figure}

Considering a mean radial velocity (RV) of $\sim$ \SI{22.5}{\kilo\meter\per\second} \citep{3D_kinematics_2021A&A...647A..19T} and that some targets might belong to multiple systems, we applied Doppler shifts corresponding to RVs ranging from \SI{10}{\kilo\meter\per\second} to \SI{45}{\kilo\meter\per\second} in steps of \SI{1}{\kilo\meter\per\second}. For stars whose best-fitting RV reached the range edges, we extended the range. Instrumental broadening corresponding to the resolution of the observed spectra and stellar rotational broadening were applied to the theoretical spectra. All the technical steps were carried out using the package PyAstronomy \citep{PyA_2019ascl.soft06010C}. The $v\sin i$ ranged from \SI{5}{\kilo\meter\per\second} to \SI{350}{\kilo\meter\per\second}, in steps of \SI{5}{\kilo\meter\per\second} initially, then ranged around best-fitting $v\sin i \pm \SI{5}{\kilo\meter\per\second}$ in steps of \SI{0.1}{\kilo\meter\per\second} to obtain more accurate values. We used Astrolib PySynphot \citep{pysynphot_2013ascl.soft03023S} to calculate the flux of the synthetic spectra corresponding to the observed wavelengths. We determined the best-fitting synthetic spectrum for each observed spectrum using the $\chi ^2$-minimization method. Two illustrative fitting results are depicted in Figure \ref{fig:vsini_demo}.

we increased our sample size to obtain a more statistically reliable result for comparing the $v\sin i$ distribution differences between disk-bearing and disk-less objects. We selected PMS stars from the catalog provided by \cite{2017A&A...599A..23V} and cross-matched them with the catalog from \citep{2009AJ....138.1116S} to classify disk-bearing and disk-less sources. The catalog of \cite{2017A&A...599A..23V} also includes stellar masses. Considering stellar masses associated with eMSTOs or split MSs are approximately from \SIrange{1.5}{5}{M_{\odot}} \citep{2018MNRAS.477.2640M}, we chose disk-bearing and disk-less sources with masses greater than \SI{1.3}{M_{\odot}} and similar mass distributions.  

We then obtained the corresponding observational spectra from European Southern Observatory (ESO) archival datasets (program ID: 094.C-0467(A)). Among the selected samples, 27 disk-bearing and 28 disk-less sources were observed by the GIRAFFE spectrograph on the Very Large Telescope with a resolution of 24000, shown as yellow dots in Fig. \ref{fig:members}. Most of the observations achieved signal-to-noise ratios (SNRs) above 50. Since PMS stars are primarily distinguished from MS stars by nuclear reactions while having similar stellar atmospheres, we employed the same procedure to calculate $v\sin i$ for the selected PMS stars as that for the MS stars (methods as mentioned in section \ref{sec:vsini_cal}), except utilizing prominent absorption lines of Fe \uppercase\expandafter{\romannumeral1} ($\sim$ \SI{6192}{\AA}).

\section{Results} \label{sec:results}
\begin{figure}
    \centering
    \includegraphics[width=0.49\textwidth]{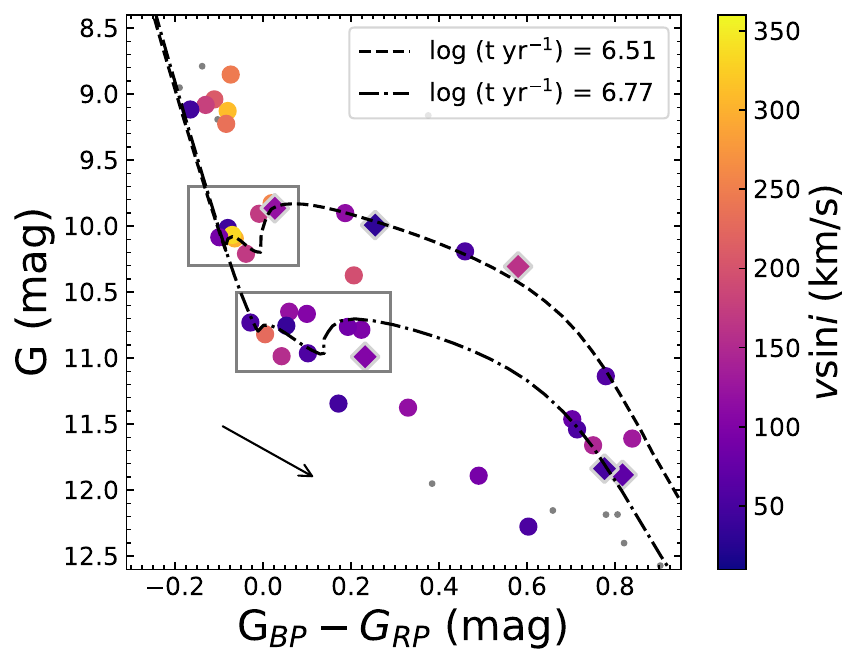}
    \caption{The CMD of NGC 2264. The best-fitting isochrones with age of $\log (t\ yr^{-1}) = 6.51$ and $\log (t\ yr^{-1}) = 6.77$ are indicated by dashed and dash-dotted lines, respectively. The black arrow represents the reddening vector, with $\Delta A_V = \SI{0.5}{mag}$.}  
    Colored circles represent stars observed by CFHT, with $v\sin i$ measurements.  The MSTOn regions are highlighted as enclosed areas by the grey boxes.  Diamonds with gray edges represent stars with disks. 
    \label{fig:vsini_pairs_2fig}
\end{figure}

NGC 2264 has an age of $\sim$ \SIrange{3}{6}{Myr}, with an age spread of $\sim$\SI{3}{Myr}, determined by the morphology of its MSTOn. As shown in Fig. \ref{fig:vsini_pairs_2fig}, two MSTOn regions appear in the CMD, shown as areas enclosed by grey rectangles. According to the best-fitting isochrones, the upper and lower MSTOn regions correspond to ages of $\sim$ \SI{3}{Myr} and $\sim$ \SI{6}{Myr}, respectively.  Noticeably, the age spread we obtained might be polluted by differential extinction. 

In order to determine whether their initial distribution of rotational velocities resembles that of most observed older open clusters, we calculated $v\sin i$ for 40 stars located within or in proximity to the MSTOn regions, employing the ESPaDOnS/CFHT spectra. Their $v\sin i$ are shown in Fig. \ref{fig:vsini_pairs_2fig} with colors. Among these stars, 10 stars are located in the lower MSTOn region, 8 stars in the upper MSTOn region, and 6 stars in the MS.  For all these stars, we utilized the same absorption line (Mg \uppercase\expandafter{\romannumeral2}, $\sim$ \SI{4481}{\AA}) to calculate their $v\sin i$. The $v\sin i$ and corresponding uncertainties are listed in Table \ref{tab:vsini_MSTOn}.  Noticeable $v\sin i$ variations appear in both MSTOn regions. The $v\sin i$ of stars in the lower MSTOn region spans from \SIrange{36.4}{228}{\kilo\meter\per\second}, and in the upper MSTOn region, it ranges from \SIrange{41.8}{333.1}{\kilo\meter\per\second}. Unlike the MSTOn stars, most MS stars (G $\sim$ \SI{9}{mag}) rotate fast, 5 out of 6 samples with $v\sin i$ larger than \SI{150}{\kilo\meter\per\second}.

\begin{deluxetable*}{clccccc} 
\caption{Information of NGC 2264 MS and MSTOn stars.}
\label{tab:vsini_MSTOn}
\tablehead{
\colhead{Stellar Location in CMD} & \colhead{Gaia EDR3 ID} & \colhead{ G$_{BP}-$G$_{RP}$ [mag]} &  \colhead{G [mag]} & \colhead{$v\sin i$ [\SI{}{\kilo\meter\per\second}]} & \colhead{$e_{v\sin i}$ [\SI{}{\kilo\meter\per\second}]}
}
\startdata
\multirow{10}{*}{\shortstack{the Lower MSTOn\\ ($\log (t\ yr^{-1}) = 6.77$)}}    &3326606484634522112&0.0529&10.7538&36.4&2.1\\
&3326683278649925888&-0.0288&10.7298&53.5&3.1\\
&3326683278649926016&0.1019&10.963&55.7&3.0\\
&3350801478522687488&0.1928&10.7626&85.8&2.5\\
&3134469748260677248&0.2235&10.7846&97.3&2.9\\
&3326684653039444480*&0.2319&10.9893&102.1&5.8\\
&3326686508465460736&0.0996&10.6641&103.7&5.4\\
&3326971144538025984&0.0594&10.6478&120.3&9.1\\
&3326702142145894784&0.0422&10.9848&154.1&9.3\\
&3326709598209096192&0.0047&10.8182&228.2&17.6\\
 \hline
\multirow{8}{*}{\shortstack{the Upper MSTOn\\ ($\log (t\ yr^{-1}) = 6.51$)}}   &3326576660381784704&-0.0801&10.0137&41.8&2.8\\
&3326941865743978880&-0.0994&10.0836&89.9&9.3\\
&3326748802670510336*&0.0267&9.8640&118.3&5.6\\
&3326695923033606912&-0.0387&10.2087&168.8&16.8\\
&3326740006577519360&-0.0090&9.9058&170.2&10.5\\
&3326736811121849216&0.0196&9.8260&256.8&17.3\\
&3326740002281694592&-0.0643&10.0926&299.8&22.2\\
&3326695854314127616&-0.0698&10.0663&333.1&33.7\\
 \hline 
 \multirow{6}{*}{\shortstack{the MS}}   &3326930771845575808&-0.0735&8.8512&245.5&18.0\\
 &3326715439364610816&-0.1107&9.0391&208.8&14.8\\
 &3350768527533815552&-0.1304&9.0812&178.0&34.5\\
 &3326717260430731648&-0.1653&9.1160&44.8&2.6\\
 &3326685924349755520&-0.0805&9.1273&310.7&18.8\\
 &3326696507149141120&-0.0838&9.2245&235.4&19.4\\
\enddata
\tablecomments{The information of MS and MSTOn stars, observed by CFHT. The lower and the upper refer to relative positions in Fig \ref{fig:vsini_pairs_2fig}. Gaia IDs with * represent disk-bearing stars. G is the Gaia G-band magnitude. G$_{BP}$ is the magnitude of Gaia blue pass band and G$_{RP}$ is the magnitude of Gaia red pass band. $e_{v\sin i}$ is the uncertainty of $v\sin i$}.
\end{deluxetable*} 

\begin{figure}
    \centering
    \includegraphics[width=0.48\textwidth]{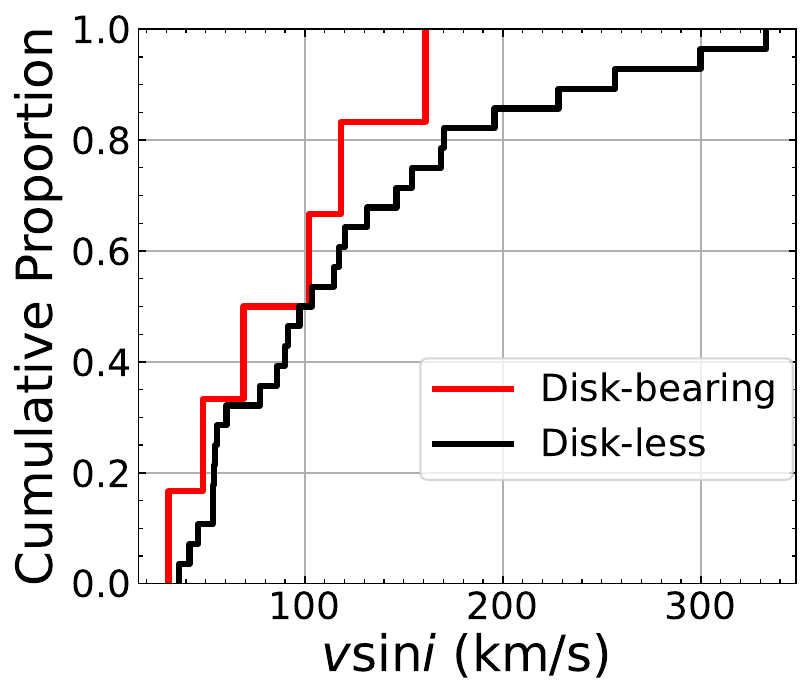}
    \caption{Cumulative $v\sin i$ proportion of PMS and MSTOn stars observed by CFHT. The disk-bearing sources were classified by cross-matching with the catalog from \citep{2009AJ....138.1116S}. The red and black lines represent disk-bearing and disk-less sources, with counts of 6 and 28, respectively.}
    \label{fig:cfht_disk_vsini}
\end{figure}

Among the stars in or near MSTOn regions, 6 PMS stars contain circumstellar disks, shown as colored diamonds with gray edges in Fig. \ref{fig:vsini_pairs_2fig}. These disk-bearing stars are identified through 
cross-matching with the catalog from \cite{2009AJ....138.1116S}. In the CFHT observed samples, there are 6 disk-bearing stars. These stars have a mass range of $\sim$\SIrange{1.9}{2.5}{M_{\odot}}, according to the best-fitting isochrones.  We derived the cumulative $v\sin i$ distribution for the disk-bearing and disk-less sources, as shown in Fig. \ref{fig:cfht_disk_vsini}. To avoid interference from rapidly rotating MS stars, we focused on the MSTOn and PMS samples in Fig. \ref{fig:cfht_disk_vsini}. Disk-bearing stars rotate on average more slowly than disk-less stars, with mean $v
\sin i$ of \SI{88.37}{\kilo\meter\per\second} and \SI{122.83}{\kilo\meter\per\second}. The $v\sin i$ of the 6 disk-bearing stars and disk-less sources that are not included in Table \ref{tab:vsini_MSTOn} are listed in Table \ref{tab:vsini_disked_cfht}. In Table \ref{tab:vsini_MSTOn}, the Gaia IDs of disk-bearing sources are marked with *. 

\begin{deluxetable*}{ccccccc} 
\caption{Information of disk-bearing and disk-less PMS stars observed by CFHT.}
\label{tab:vsini_disked_cfht}
\tablehead{
\colhead{Gaia EDR3 ID} & \colhead{ G$_{BP}-$G$_{RP}$ [mag]} & \colhead{G [mag]} & \colhead{$v\sin i$ [\SI{}{\kilo\meter\per\second}]} & \colhead{$e_{v\sin i}$ [\SI{}{\kilo\meter\per\second}]} & \colhead{Disk-bearing?}
}
\startdata
3326717226070995328&0.2548&9.9916&31.2&4.4&Yes\\
3326691009590655616&0.7760&11.8368&48.6&1.9&Yes\\
3326712411411788032&0.8169&11.8842&69.0&2.4&Yes\\
3326684653039444480&0.2319&10.9893&102.1&5.8&Yes\\
3326748802670510336&0.0267&9.8640&118.3&5.6&Yes\\
3326741277887838080&0.5798&10.3053&161.0&7.5&Yes\\
\hline
3326589025591649408&0.1714&11.3438&46.1&1.4&No\\
3326678193408399232&0.4591&10.1906&53.6&3.3&No\\
3326816422637041920&0.603&12.2757&54.2&4.9&No\\
3326689807000188032&0.7137&11.5395&54.8&1.7&No\\
3326703619614642560&0.779&11.1363&60.5&2.9&No\\
3326540376498237312&0.7027&11.4624&77.3&2.0&No\\
3350770795276039168&0.49&11.8902&91.4&3.0&No\\
3326737051640013952&0.1868&9.9008&114.7&7.5&No\\
3326740998714107008&0.3295&11.3746&117.4&3.4&No\\
3326905478782410368&0.8391&11.6089&131.6&3.2&No\\
3326682552799138176&0.7501&11.6597&146.3&3.0&No\\
3326736811121849344&0.2063&10.3725&195.8&11.4&No\\
\enddata
\tablecomments{The information of disk-bearing stars and disk-less PMS stars, observed by CFHT.  G is the Gaia G-band magnitude. G$_{BP}$ is the magnitude of Gaia blue pass band and G$_{RP}$ is the magnitude of Gaia red pass band. $e_{v\sin i}$ is the uncertainty of $v\sin i$}.
\end{deluxetable*}

\begin{figure}
    \centering
    \includegraphics[width=0.47\textwidth]{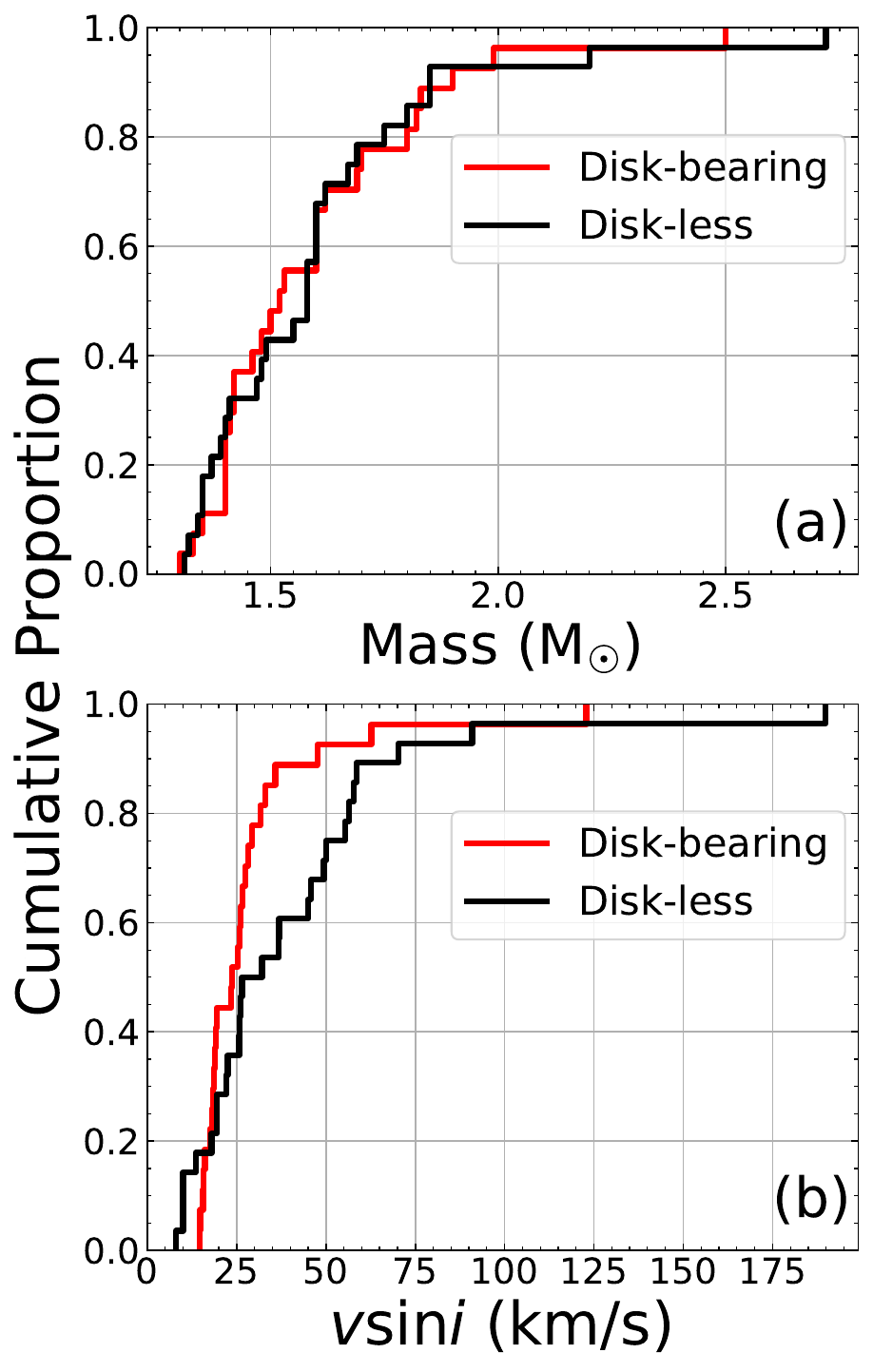}
    \caption{
    Cumulative mass and $v\sin i$ proportion of selected samples from the catalog provided by \cite{2017A&A...599A..23V}, which includes stellar masses. Red and black lines represent disk-bearing and disk-less sources, respectively. The disk-bearing and disk-less sources were classified by cross-matching with the catalog from \citep{2009AJ....138.1116S}. 
    }
    \label{fig:mass_vsini_2fig}
\end{figure}

The 55 PMS samples, selected from the catalog by \cite{2017A&A...599A..23V} and classified as disk-bearing and disk-less sources by cross-matching with the catalog from \cite{2009AJ....138.1116S}, provided more statistically reliable results.  The masses of the 27 disk-bearing and 28 disk-less stars range from $\sim$ \SIrange{1.3}{2.8}{M_{\odot}}, as shown in Fig. \ref{fig:mass_vsini_2fig}(a). Their $v\sin i$ are listed in Table \ref{tab:vsini_ESO_disked)} and \ref{tab:vsini_ESO_diskless)}. The mean velocities of disk-bearing and disk-less stars are \SI{28.8}{\kilo\meter\per\second} and \SI{40.4}{\kilo\meter\per\second}. Fig. \ref{fig:mass_vsini_2fig}(b) displays the measured  $v\sin i$ cumulative proportions of disk-bearing and disk-less sources. We immediately found that although the mass distributions between the disk-less and disk-bearing stars do not show systematic differences, their $v \sin i$ distributions are obviously different. We applied a two-sample Kolmogorov-Smirnov (K-S) test to the disk-bearing and disk-less samples, yielding a $p$ value of 0.04, which indicates that they harbor different distributions. The disk-bearing samples are on average lower than disk-less samples, especially in the $v \sin i$ range of \SIrange{25}{75}{\kilo\meter\per\second}.  The $v\sin i$ of disk-bearing stars mainly concentrate on $v\sin i < \sim$ \SI{35}{\kilo\meter\per\second}, while the $v\sin i$ of disk-less sources spans a larger range.

\section{Discussion} \label{sec:discussion}

We first discuss how our results can examine the disk-locking and binary-merger models, as proposed by \cite{2020MNRAS.495.1978B,2022NatAs...6..480W}. Our observational results generally align with the disk-locking model. Our results suggest that stars that are able to sustain their disks for longer periods tend to have smaller rather than larger rotational velocities. Even though we can not determine whether the $v\sin i$ cumulative proportions of disk-bearing and disk-less stars share the same distribution or not, due to the small sample of disk-bearing stars observed by CFHT (a count of 6), we noticed that the disk-bearing stars lack of fast rotators with $v\sin i$ over \SI{170}{\kilo\meter\per\second}, which cannot be explained by measurement uncertainties.  The $v\sin i$ of the fastest disk-bearing rotator is \SI{161}{\kilo\meter\per\second}. 7 Stars with $v\sin i$ larger than \SI{161}{\kilo\meter\per\second} are found in the disk-less samples with a count of 28.
Assuming random inclination angles of the spin axis, this may imply that in their mass range ($\sim$ \SIrange{1.9}{3}{M_{\odot}}), disk regulation plays an important role in stellar angular momentum distribution, Additionally, since these stars are close to the end of PMS stages, their rotation rates are unlikely to change significantly due to contraction. This indicates disks constrain the upper limit of stellar rotation rates, resulting in slow rotators. Stars lose their disks in early evolution stages would have a higher probability of evolving to fast rotators than disk-bearing stars. If so, they would form a bimodal distribution of surface rotational velocity as they transition into MS. It would help testify to this process by simulating a realistic cluster and tracking their statistical rotation rate evolution. Due to the small sample size, we can not conclude the $v\sin i$ difference between disk-bearing and disk-less stars as a statistically reliable result. A larger sample size would enhance the reliability of our conclusions.



With a larger sample size that includes 27 disk-bearing and 28 disk-less stars selected from the catalog provided by \cite{2017A&A...599A..23V} and classified from the catalog of \cite{2009AJ....138.1116S}, PMS stars with circumstellar disks exhibit, on average, lower $v\sin i$, as shown in Fig. \ref{fig:mass_vsini_2fig}(b). The difference is unlikely mainly caused by different inclination angles.  The $v\sin i$ difference is prominent in \SIrange{25}{75}{\kilo\meter\per\second}, which could not be produced by the same $v_{eq}$ distributions multiplied by $\sin i$ with random inclination probabilities.  

In the model proposed by \cite{2020MNRAS.495.1978B}, they set initial surface velocities to $\sim$ \SIrange{20}{70}{\kilo\meter\per\second} for their model PMS stars. Surface velocities of these stars then evolve to $\sim$ \SIrange{60}{230}{\kilo\meter\per\second} at the arrival of the MS and maintain in the MS.  The $v\sin i$ differences of our observational disk-bearing and disk-less PMS stars indeed dominant in the \SIrange{25}{75}{\kilo\meter\per\second} range. The $v\sin i$ could potentially enlarge to $\sim$ 3 times as they reach the MS. Notably, it aligns with the observation that MSTOn stars exhibit a broad distribution of $v\sin i$, ranging from $\sim \SIrange{50}{300}{\kilo\meter\per\second}$, and the $v\sin i$ difference is larger in the samples close to PMS ending stages, as shown in Fig. \ref{fig:cfht_disk_vsini}.

However, the binary-merger model would foresee a different outcome \citep{2022NatAs...6..480W}. If the highly active binary mergers during the star-forming stage are primarily driven by the ejection of the circumbinary disk, then it would suggest that most disk-less stars are slow rotators. Conversely, it is well known that close stellar encounters can significantly alter the morphological characteristics of circu-stellar disks, leading to the loss of disk mass, a reduction in their radius, or even complete destruction \citep[e.g.,][]{2023MNRAS.520.6159C,2018ApJ...868....1V,2019MNRAS.482..732C}.  
Based on the work of \cite{2022NatAs...6..480W}, binary mergers produce slow rotators by generating strong magnetic fields. Additionally, triple and higher-order multiple systems tend to foster binary mergers. Considering both disk ejection and interactions with external stars can lead to binary merger, as well as disrupt disks, if binary mergers primarily arise from the intense multi-body interactions within the dense environments of star clusters, it can also be anticipated that, within the framework of binary merger, disk-less stars are likely to be slow rotators. This stands in contrast to our observations, however. Additionally, if binary mergers occur at a frequency of $\sim$25\% to 50\% (slow rotators frequency) in star-forming regions, these events would be worth to track, potentially yielding several occurrences each year in the Milky Way.

\begin{figure}
    \centering
    \includegraphics[width=0.47\textwidth]{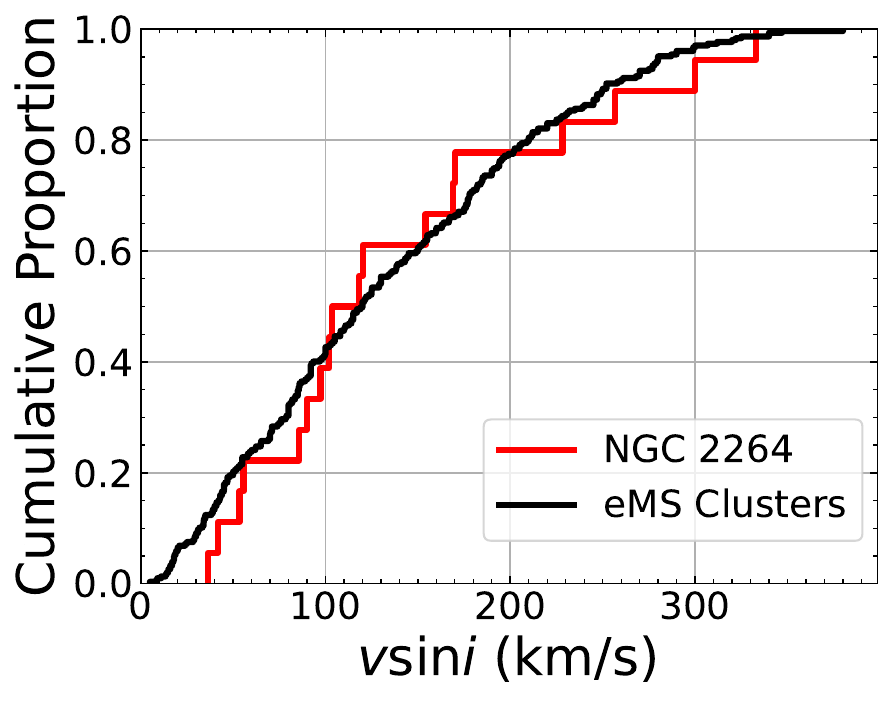}
    \caption{The $v\sin i$ distributions of NGC 2264 MSTOn stars and 7 evolved young open clusters displaying eMSs (combined of NGC 1818  \citep{2018AJ....156..116M}, NGC 2422 \citep{2022ApJ...938...42H}, NGC 2287 \citep{2019ApJ...883..182S}, NGC 6705  \citep{2018ApJ...863L..33M}, NGC 2818 \citep{2018MNRAS.480.3739B}, NGC 5822  \citep{2019ApJ...876..113S}, NGC 2423  \citep{2024ApJ...968...22B}}), represented by red and black lines, respectively.
    \label{fig:2423}
\end{figure}

At the ends of PMS and beginning of MS stages,  MSTOn stars of NGC 2264 show wide stellar rotation distribution, as shown in Fig. \ref{fig:vsini_pairs_2fig} and Table \ref{tab:vsini_MSTOn}. According to the best-fitting isochrones, The masses of the MSTOn stars range from \SIrange{2}{3}{M_{\odot}}, which falls within the typical mass range of eMS stars ($\sim$\SIrange{1.5}{5}{M_{\odot}}). Since $\sin i$ of these stars can not be determined and \cite{2020MNRAS.495.1978B}'s model suggests that differences of surface velocities would persist in the MS stage,  we compared the $v\sin i$ distributions of NGC 2264 MSTOn stars with those of evolved young open clusters displaying eMSs. Specifically, we combined the $v\sin i$ distributions of NGC 1818 ($\sim$ \SI{40}{Myr}, \citep{2018AJ....156..116M}), NGC 2422 ($\sim$ \SI{90}{Myr}, \citep{2022ApJ...938...42H}), NGC 2287 ($\sim$ \SI{200}{Myr} \citep{2019ApJ...883..182S}), NGC 6705 ($\sim$ \SI{300}{Myr}, \citep{2018ApJ...863L..33M}), NGC 2818 ($\sim$ \SI{800}{Myr} \citep{2018MNRAS.480.3739B}), NGC 5822 ($\sim$ \SI{900}{Myr}  \citep{2019ApJ...876..113S}), NGC 2423 ($\sim$ \SI{1}{Gyr} \citep{2024ApJ...968...22B}) together and compare it with the $v\sin i$ distribution of NGC 2264 MSTOn stars. The available masses are relatively close to MSTOn stars in NGC 2264, with NGC 2422 $\sim \SI{2}{M_\odot}$ \citep{2022ApJ...938...42H}, NGC 2287 $\sim \SIrange{3}{4}{M_\odot}$ \citep{2019ApJ...883..182S}, NGC 2818 $\sim \SIrange{1.3}{2}{M_\odot}$ \citep{2018MNRAS.480.3739B}, NGC 2423 $\sim \SIrange{1.2}{2}{M_\odot}$ \citep{2024ApJ...968...22B}. The distributions are very similar, as shown in Fig. \ref{fig:2423}.  We also run an Anderson–Darling test for k-samples to test the hypothesis that the NGC 2264 MSTOn stars and the eMS stars harbors the same distribution. The test result shows a significance level larger than 0.25, indicating that it is insufficient to reject the hypothesis. In this case, we suggest that the observed wide stellar rotation distribution in many young clusters can occur in their early stages and may persist during MS stages. 

Furthermore, if the observed luminosity spread is not driven by the effects of star-spots or disk self-distinguish \citep[e.g.][]{2009A&A...504..461F,2021ApJ...908...49F,2013ApJS..207....5F,2013A&A...549A..15F}, then the cluster may have an genuine age spread of $\sim$ \SIrange{3}{6}{Myr}. If the slowly rotating MS stars are dynamically tidally-locked, they would harbor very short orbital periods. Based on the synchronisation timescale formula provided by \cite{Zahn_1975A&A....41..329Z}, assuming stellar masses of \SIrange{2}{5}{M_{\odot}}, mass ratio of 0.4, and synchronization time of \SI{6}{Myr}, the derived binary separations are $\sim$ \SIrange{0.030}{0.065}{AU}, corresponding to periods of \SIrange{1.12}{2.27}{d}\footnote{In our calculation, we applied the empirical mass-radius relation from \cite{1991Ap&SS.181..313D} to estimate stellar radius}. For a MS star with the solar radius, a period of \SI{2}{d} corresponds to a equatorial velocity of $\sim$ \SI{25}{\kilo\meter\per\second}, classifying the star as a slow rotator. However, these short-period binaries are not common in observation. For stars with masses of \SIrange{2}{5}{M_{\odot}}, the binary frequency with periods of \SIrange{3}{22}{d} is only $7\pm 2\%$ \citep{2017ApJS..230...15M},  while binaries with periods of $\sim$ \SIrange{1}{3}{d} occur at an even lower frequency (see Fig. 31 in the article by \cite{2017ApJS..230...15M}). This indicates that the stellar rotational velocity distribution might not be mainly regulated by tidal interactions for NGC 2264. In addition, even if a high frequency of short-period binaries appears in star-forming regions, these would need to be very close binaries. In such cases, the companion star might form within the disk, potentially resulting in slow-rotating, disk-less sources. However, we note that the tidal interactions might slow down the rotation of stars more efficiently than that expected by dynamical tides theory, as the orbital circularization timescale of some binaries are found to be shorter than the expectation of the theory \citep{2008EAS....29...67Z,2015MNRAS.453.2637D}. Thus, the contribution of tidal interactions can not be ruled out. In addition, these stars might be synchronized at birth. Searching for synchronized PMS stars or spectroscopic PMS binaries could help to verify this hypothesis. 

In some young clusters, low mass ($<\sim$\SI{2}{M_{\odot}}) PMS stars display bimodal period distributions, and disk-bearing objects rotate more slowly on average than sources without circumstellar disks, indicating magnetic star-disk interaction are sufficient to slow low-mass stars down, such as NGC 2264 \citep{2007ApJ...671..605C,2017A&A...599A..23V} and IC 348\citep{2006ApJ...649..862C}. However, only 18 MSTOn stars are found in NGC 2264, we can not determine whether their rotation distribution is bimodal or not. Further observation of more massive clusters might help address this issue. 

MS stars are more massive than MSTOn stars, their masses are between 4 to 5 \SI{}{M_{\odot}}. Unlike MSTOn stars, most MS stars are fast rotators ($v\sin i > \SI{150}{\kilo\meter\per\second})$, as shown in Fig. \ref{fig:vsini_pairs_2fig} and Table \ref{tab:vsini_MSTOn}, indicating that magnetic star-disk interaction might not be sufficient to slow them down.  The result aligns with the model developed by  \cite{2012ApJ...748...97R}, where they found that massive stars (e.g., $M_{\rm star} > \SI{6}{M_{\odot}}$ ) are preferentially rapid rotators. They suggested that due to greater inertia and relatively weaker magnetic fields, the rotation of massive stars was more difficult to slow down, compared to low-mass stars, unless their disk lifetimes are much longer or magnetic fields are much stronger than expected. It is not clear whether $v\sin i$ of massive stars is dependable on their birth environment \citep{2012ApJ...748...97R}. The maximum stellar mass at which disk-locking can work effectively is expected; however, it is unclear whether this limit depends on the environment of the star cluster and requires future investigations.

\section{Conclusion} \label{sec:conclusion}
In this work, we investigated a star forming cluster, NGC 2264, to examine whether disk-locking plays an important role in setting stellar rotational distributions, which causes eMSs observed in most young open clusters.

Our main results can be summarized as follows.

\begin{itemize}
    \item[1.] Disks regulate stellar rotational velocity distributions of intermediate-mass PMS stars, which would evolve to stars causing eMSs. Disk-less stars rotate on average faster than disk-bearing stars.
    \item[2.] Intermediate-mass MSTOn stars of NGC 2264 show similar rotational velocity distributions with evolved young open clusters displaying eMSs. The distribution could appear in very early stages and may persist during the MS, resulting eMSs.
    \item[3.]  Magnetic star-disk interaction might not be sufficient to regulate the angular momenta of massive stars ($M_{\rm star} > \sim\SI{4}{M_{\odot}}$ for NGC 2264).
\end{itemize}

Therefore, we conclude that at least in the star cluster NGC 2264, disk-locking has an important role in regulating stellar rotation. Future studies on more star-forming regions in the local group will shed light on whether this is a common phenomenon, helping us understand how interactions between stellar disks and stars early stage would affect the later evolution of stellar populations.

\begin{acknowledgments}

This work was supported by the National Natural Science Foundation of China (NSFC) through grant 12233013 and 12073090. MF acknowledges financial supports by the International Partnership Program of Chinese Academy of Sciences (Grant number 019GJHZ2023016FN). This work has made use of data from Canada\text{-}France\text{-}Hawaii Telescope (CFHT). The data were obtained through the Telescope Access Program (TAP), which has been funded by the TAP member institutes. This work also used data from the European Space Agency (ESA) mission {\it Gaia}(\url{https://www.cosmos.esa.int/gaia}), processed by the {\it Gaia} Data Processing and Analysis Consortium (DPAC, \url{https://www.cosmos.esa.int/web/gaia/dpac/consortium}). Funding for the DPAC has been provided by national institutions, in particular the institutions participating in the {\it Gaia} Multilateral Agreement. This work was also based on data obtained from the ESO Science Archive Facility with DOI: \url{https://doi.org/10.18727/archive/27}.

\end{acknowledgments}

\software{MIST \citep{2016ApJ...823..102C}, Astropy \citep{Price-Whelan_2018}, Matplotlib \citep{2007CSE.....9...90H}, SciPy \citep{2020NatMe..17..261V}, PyAstronomy \citep{2019ascl.soft06010C}, Astrolib PySynphot \citep{2013ascl.soft03023S}, TOPCAT \citep{2005ASPC..347...29T}.
}

%




\clearpage
\appendix
\section{Membership Comparison with Previous Work}
\begin{figure}
    \centering
    \includegraphics[width=1.0\textwidth]{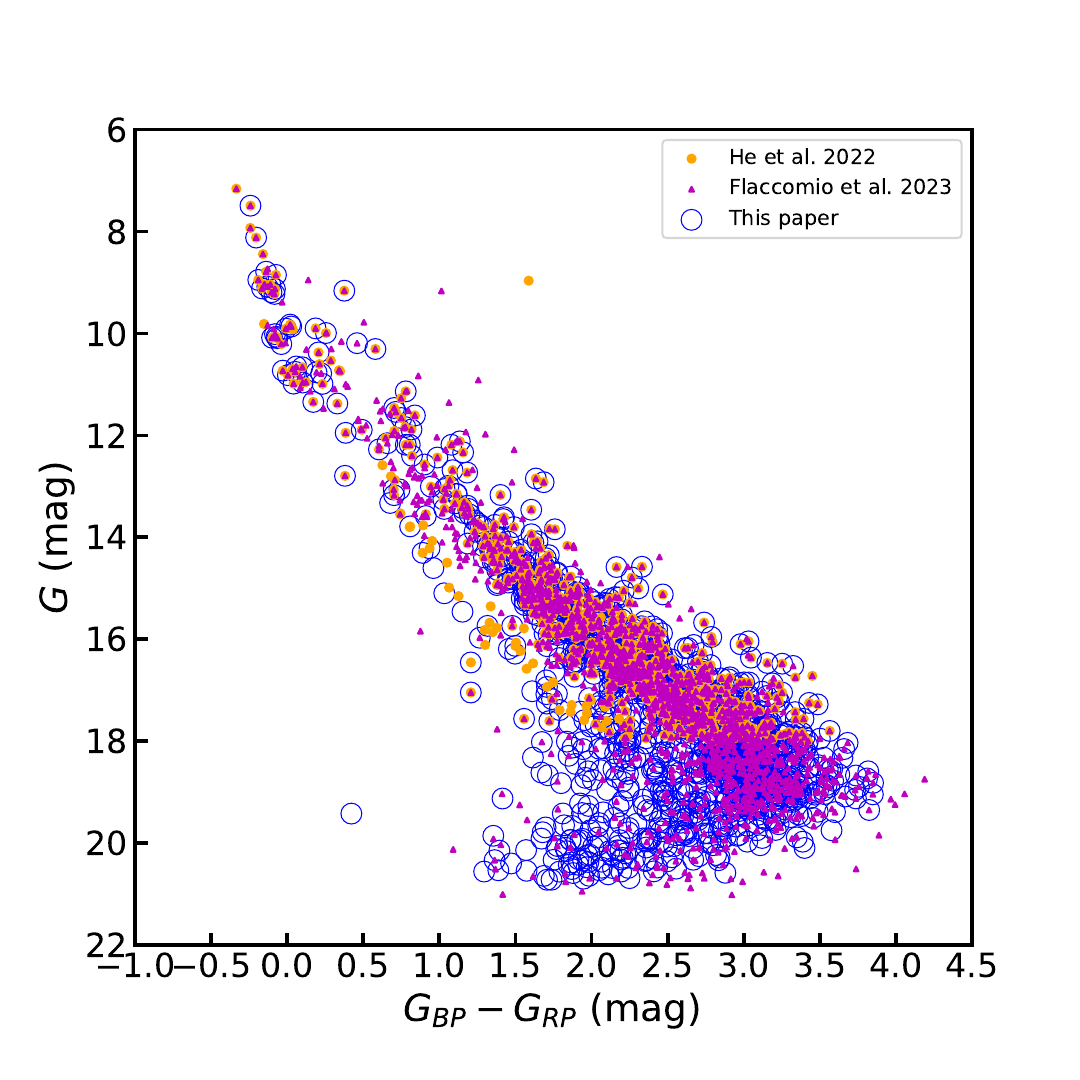}
    \caption{The CMDs of the member stars selected in this paper (open blue circles), from \cite{2022ApJS..262....7H} (orange solid dots), and from \cite{2023A&A...670A..37F} (magenta triangles). The member stars from \cite{2023A&A...670A..37F} are those compromise samples between contamination and sample completeness suggested by \cite{2023A&A...670A..37F} (for details, see \cite{2023A&A...670A..37F}) who were also cross-matched with Gaia EDR3.}
    \label{fig:comparison}
\end{figure}

\section{$v\sin i$ of NGC 2264 PMS stars}

\begin{deluxetable*}{ccccccc} 
\caption{Information of NGC 2264 disk-bearing PMS stars with $v\sin i$ measurement}
\label{tab:vsini_ESO_disked)}
\tablehead{
\colhead{Gaia EDR3 ID} & \colhead{ G$_{BP}-$G$_{RP}$ [mag]} & \colhead{G [mag]} & \colhead{$v\sin i$ [\SI{}{\kilo\meter\per\second}]} & \colhead{$e_{v\sin i}$ [\SI{}{\kilo\meter\per\second}]}
}
\startdata
3326712278268676736&1.5811&14.9069&14.7&0.7\\
3326698736235918592&1.4247&13.6072&14.8&0.7\\
3326695854314131200&1.634&12.8465&15.6&0.4\\
3326723376464152960&1.4534&14.8021&15.7&1.4\\
3326928705964373888&1.5663&14.5454&16.1&1.0\\
3326644245987098624&1.4198&13.6424&17.5&1.2\\
3326698186480571264&1.1649&13.5257&18.1&0.7\\
3326717294790476672&1.9107&14.979&18.4&2.4\\
3326702863700390144&1.4938&14.3063&18.6&4.2\\
3326685099716035456&1.715&14.8571&19.0&1.1\\
3326685615112100992&1.3953&14.5218&19.1&1.5\\
3326708807935122560&1.5779&14.852&19.5&3.8\\
3326710663360978816&1.402&14.0083&23.5&0.6\\
3326905272623980288&1.1897&13.4423&23.7&0.8\\
3326686572888883072&1.5083&14.4109&25.3&0.9\\
3326739628620404352&1.473&14.3629&25.9&0.9\\
3326905036401556480&1.8735&15.599&26.1&4.3\\
3326697228703282432&1.1127&13.1558&26.6&0.8\\
3326717019912569088&1.49&13.799&27.4&0.6\\
3326689738279228800&1.4688&14.2774&28.2&1.1\\
3326698568733398912&1.2909&13.7984&29.4&0.7\\
3326689291604121216&1.3855&14.1313&31.6&1.1\\
3326691009590655616&0.776&11.8368&33.0&0.8\\
3326696060472549120&2.3315&14.5783&35.8&4.0\\
3326712411411788032&0.8169&11.8842&47.6&0.9\\
3326697159983811712&2.1528&15.9091&62.7&2.8\\
3326741101798707584&1.1346&12.1178&122.9&0.4\\
\enddata
\tablecomments{The information of 27 disk-bearing stars, observed by VLT. G is the Gaia G-band magnitude. G$_{BP}$ is the magnitude of Gaia blue pass band and G$_{RP}$ is the magnitude of Gaia red pass band. $e_{v\sin i}$ is the uncertainty of $v\sin i$}.
\end{deluxetable*}

\begin{deluxetable*}{ccccccc} 
\caption{Information of NGC 2264 disk-less stars with $v\sin i$ measurement}
\label{tab:vsini_ESO_diskless)}
\tablehead{
\colhead{Gaia EDR3 ID} & \colhead{ G$_{BP}-$G$_{RP}$ [mag]} & \colhead{G [mag]} & \colhead{$v\sin i$ [\SI{}{\kilo\meter\per\second}]} & \colhead{$e_{v\sin i}$ [\SI{}{\kilo\meter\per\second}]}
}
\startdata
3326904521005483136&1.3659&13.9727&8.0&0.9\\
3326724647774743808&1.5218&14.9638&9.2&1.5\\
3326933073948600064&1.1624&13.3101&10.0&3.7\\
3326697400501968384&0.7977&12.9221&10.0&3.7\\
3326739933562218496&1.0866&12.6832&10.0&2.0\\
3326684648743194112&1.104&12.3089&10.0&1.4\\
3326711487994693504&1.0261&13.2572&13.5&0.8\\
3326698289559319936&1.0695&12.8777&17.6&0.5\\
3326685099716035456&1.715&14.8571&19.0&1.1\\
3326714958328277376&1.5073&14.7479&19.5&1.1\\
3326741346607877248&2.1066&15.2819&19.5&3.0\\
3326739903498311808&1.5375&14.2463&22.1&1.0\\
3326713102902390656&1.3609&14.2609&22.5&0.6\\
3326704852270253056&0.944&13.0054&22.6&0.7\\
3326929427518677504&1.1841&13.3804&25.3&0.6\\
3326895724912475648&1.1759&13.5883&25.8&1.1\\
3326741342310109056&0.8056&12.1843&26.2&0.6\\
3326685301578222080&0.9242&13.2109&26.5&0.9\\
3326739869138571904&1.1563&12.3344&32.1&0.4\\
3326685370298017408&0.9871&12.4352&36.8&0.6\\
3326931699558507648&1.365&13.7555&36.9&0.7\\
3326907059330544640&1.6645&15.0257&42.8&7.5\\
3326689768344207488&0.9027&12.5715&45.0&1.1\\
3326696816386428800&1.3269&14.1331&45.8&1.1\\
3326703512240456064&1.1137&13.1467&49.4&1.3\\
3326689807000188032&0.7137&11.5395&50.0&1.1\\
3326712793665773184&1.2194&13.7005&55.3&1.2\\
3326712999823177088&0.6432&12.4573&56.5&1.0\\
3326737665818107008&1.286&13.0662&57.7&2.9\\
3326716637659605760&1.1157&11.5506&58.6&0.4\\
3326693346053390848&1.2998&13.9789&70.3&1.6\\
3326708356963565824&1.0341&13.0527&90.9&2.4\\
3326929225657356288&1.1481&13.4211&189.9&5.5\\
\enddata
\tablecomments{The information of 28 disk-less stars, observed by VLT. G is the Gaia G-band magnitude. G$_{BP}$ is the magnitude of Gaia blue pass band and G$_{RP}$ is the magnitude of Gaia red pass band. $e_{v\sin i}$ is the uncertainty of $v\sin i$}.
\end{deluxetable*}

\bibliography{sample631}{}
\bibliographystyle{aasjournal}



\end{CJK*}
\end{document}